\documentclass[reprint,aps, superscriptaddress,nofootinbib]{revtex4-2}

\usepackage{dcolumn}% Align table columns on decimal point
\usepackage{bm}% bold math

\usepackage{amsmath}
\usepackage{amssymb}
\usepackage{amsfonts}
\usepackage{amsthm}
\usepackage{color}
\usepackage{subfig}
\usepackage{MnSymbol}
\usepackage{graphicx}
\usepackage{float}

\begin{document}

\title{Dynamics of the order parameter in symmetry breaking phase transitions}

\date{\today}

\author{Fumika Suzuki}
\email{fsuzuki@lanl.gov}
 \affiliation{%
Theoretical Division, Los Alamos National Laboratory, Los Alamos, New Mexico 87545, USA
}
 \affiliation{%
Center for Nonlinear Studies, Los Alamos National Laboratory, Los Alamos, New Mexico 87545, USA
}

\author{Wojciech H. Zurek}
%\email{whz@lanl.gov}
 \affiliation{%
Theoretical Division, Los Alamos National Laboratory, Los Alamos, New Mexico 87545, USA
}

  \begin{abstract}
The formation of topological defects in second-order phase transitions can be investigated by solving partial differential equations for the evolution of the order parameter in space and time, such as the Langevin equation. We demonstrate that the ordinary differential equations governing either the temporal or spatial dependence in the Langevin equation provide surprisingly substantial insights into the dynamics of the phase transition. The temporal evolution of the order parameter predicts the essence of the adiabatic-impulse scenario, including the scaling of the freeze-out time $\hat{t}$, which is crucial to the Kibble-Zurek mechanism (KZM). In particular, Bernoulli differential equations that arise in the overdamped case can be solved analytically. The spatial part of the evolution, in turn, leads to the characteristic size of domains that choose the same broken symmetry. Apart from the fundamental insights into the KZM, this finding enables the exploration of Kibble-Zurek scaling using ordinary differential equations over a large range of quench timescales, which would otherwise be difficult to achieve with numerical simulations of the full partial differential equations.
 \end{abstract}

\maketitle

%Density of topological defects left behind second order phase transitions scales as a power law of the quench time. The exponent is set by the critical exponents of that phase transition. This scaling was deduced via the intuitive argument \cite{whz} that relies on critical slowing down: The size of the domains that determine the density of defects is set when the relaxation timescale becomes comparable to the time interval before or after the critical point is traversed. This heuristic argument leads to surprisingly successful predictions that have bben confirmed by numerical simulations \cite{ } and by experiments \cite{ }. Here we look ``under the hood'' of that scaling by separately exploring time and space dependence of the order parameter in course of the quench. 

The Kibble-Zurek mechanism (KZM) combines the inevitability of topological defect formation in cosmological phase transitions noted by Kibble \cite{kibble,kibble2} with the theory \cite{whz,whz2,whz3,kzm} that relates their density to the critical slowing down and, hence, to the universality class of the second-order phase transition.
%The Kibble-Zurek mechanism (KZM) \cite{kibble,kibble2,whz,whz2,whz3,kzm} predicts topological defect density as a function of the quench timescale in second-order phase transitions. 
It finds applications 
%across a wide range of fields, including 
in condensed matter physics \cite{exp, dorner,  bec,bec2,reichhardt, reichhardt2,ralf,adolfo,anderson, jacek, mithun, bayocboc,wit,carmi,monaco,sadler,weiler,chiara,benni,ulm,pyka,chae,lin,griffin,dona,mono,choma,navon,rysti,saito,cyn}, cosmology \cite{kibble,kibble2,whz,whz2,whz3,cosmology,cosmology2,cosmology3,cosmology4}, chemistry \cite{nik,nik2}, as well as quantum simulation and quantum computing \cite{damski,zoller,dziarmaga,dziarmaga2, polkovnikov,polkovnikov2,lukasz, qc,qc2,qc3,qc4, bando,gardas,universal,anders}.

\begin{figure}
{%
\includegraphics[clip,width=0.9\columnwidth]{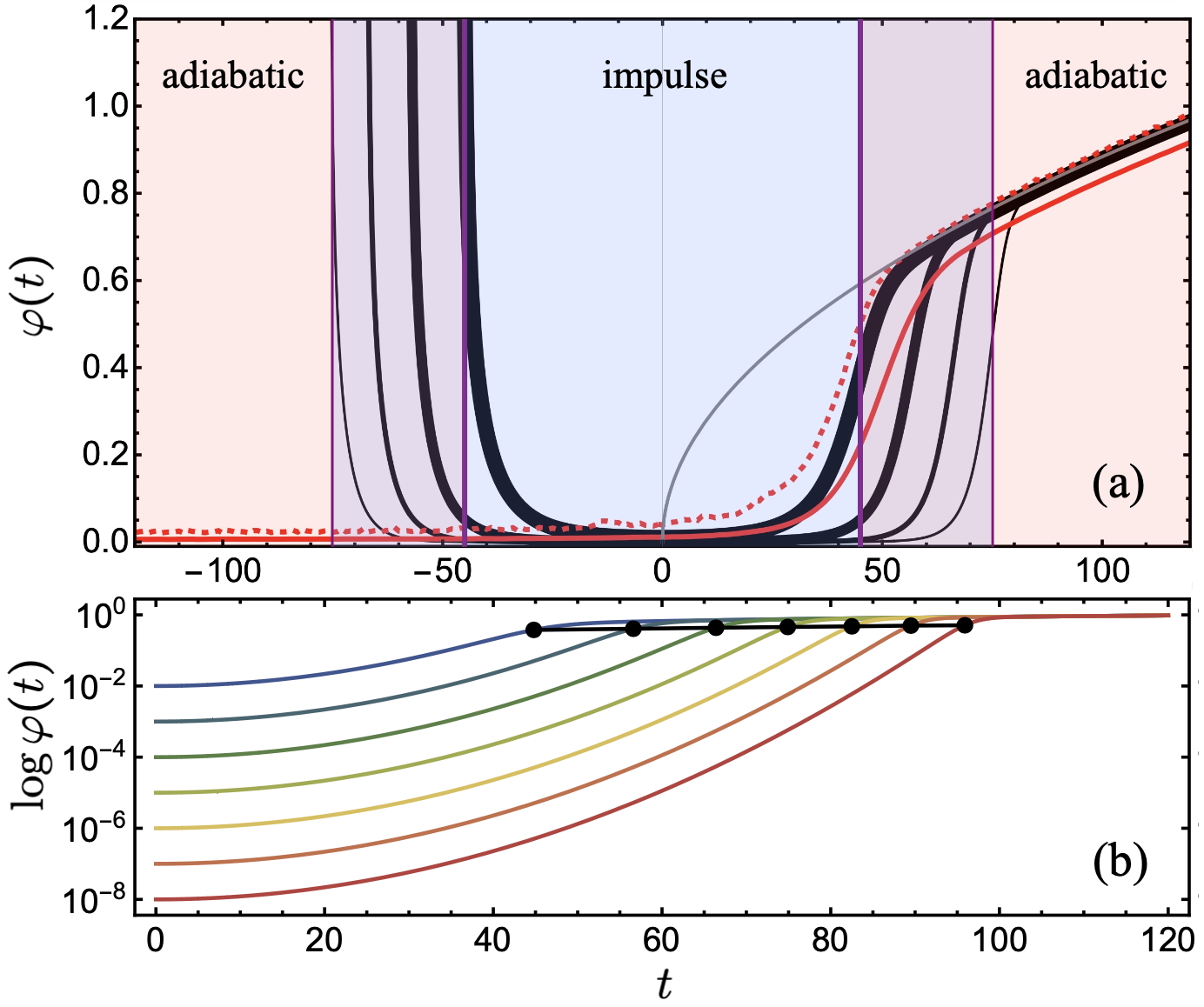} 
}
\caption{(a) The order parameter $\varphi (t)$, Eq. (\ref{solu}), for $\eta=1$, $\tau_Q=128$, and $\varphi (0)=10^{-2}, 10^{-3}, 10^{-4}, 10^{-5}$, from thick to thin line, as well as the numerical solutions (see Fig. \ref{fig2}). The  grey line is equilibrium $|\varphi_{\rm min}|=\sqrt{\epsilon}$. The thick and thin purple vertical lines indicate $\pm \hat{t}$ for $\varphi (0)=10^{-2}$ and $10^{-5}$ respectively. The numerical results $\sqrt{\langle \Phi (x,t)^2\rangle}$ (solid red), $\displaystyle\mbox{max}_x |\Phi (x,t)|$ (dashed red) are obtained from  Eq. (\ref{langevin}) with $\eta=1$ and $\theta=10^{-4}$. 
%$|\varphi_{\rm min}|
%the location of the minimum of the potential $V$. 
(b) Plots of $\log \varphi(t)$ for $\varphi (0)=10^{-2} ...10^{-8}$. 
%Note the logarithmic insensitivity to $\varphi (0)$.
%$10^{-3}$, $10^{-4}$, $10^{-5}$, $10^{-6}$, $10^{-7}$, 
%the order parameter evolves as 
As $\varphi (t) \approx {\varphi (0)} e^{t^2/4\eta\tau_Q}$ when $t \in [-\hat{t},+\hat{t}]$, perturbations present at $-\hat t$ reappear at $+\hat t$, so they are in effect ``frozen''. 
Noise (see Eq. (1)) added when $t \in [-\hat{t},+\hat{t}]$ is amplified, but the freeze-out time $+\hat{t}$ is insensitive (depends logarithmically) on $\varphi(0)$. 
}
\label{fig1}
\end{figure}

The key insight \cite{whz} that leads to KZM scaling is the realization that, near the critical point of the second-order phase transition, critical slowing down will result in a time interval $[-\hat{t},\hat{t}]$ where the order parameter is too sluggish to adjust to the potential that is changing faster than its reaction time. Thus, while outside this interval the order parameter can be in approximate equilibrium, within the interval $[-\hat{t},\hat{t}]$ its evolution
``cannot keep up''. Fluctuations imparted after $-\hat t$ seed topological defects that germinate after $+\hat t$ in ways that depend on the nature of the system \cite{sonic}. KZM appears to be fairly insensitive to the details of that evolution. 
%For instance, the resulting scalings are the same whether or not the order parameter is conserved. 
%along with its near independence of the spatial evolution generator 
This broad applicability of KZM suggests that the critical slowing down is key to its success. We focus first on the temporal evolution of the order parameter (Fig. \ref{fig1}). We also discuss how the spatial structure  responds to the quench-induced transition by imprinting the symmetry-breaking domains.

The numerical study of topological defect formation involves simulations of the Langevin equation with a time-dependent potential. For example, one  solves a Langevin equation with  the time-dependent Landau-Ginzburg potential,  $V(\Phi)=(\Phi^{4}-2\epsilon (t) \Phi^{2})/8$,  for the real scalar field $\Phi$ representing the order parameter \cite{lg,lg2,yates,antunes, jacek2,fumika}:
\begin{eqnarray}\label{langevin}
\ddot{\Phi} (\mathbf{x},t)+\eta \dot{\Phi}(\mathbf{x},t) -\nabla^2 \Phi (\mathbf{x},t) +\partial_{\Phi} V(\Phi) =\vartheta (\mathbf{x},t)
\end{eqnarray}
where  the noise term $\vartheta$ has correlation properties,
\begin{eqnarray}
\langle \vartheta (\mathbf{x},t), \vartheta  (\mathbf{x}',t')\rangle = 2\eta \theta \delta (\mathbf{x}'-\mathbf{x})\delta (t'-t).
\end{eqnarray}
Here, $\eta$ represents the damping constant, and $\theta$ is the temperature of the reservoir. Dimensionless distance from the critical point $\epsilon$ represents a quench, $\epsilon(t) = t/\tau_Q$ with $\tau_Q$ a quench timescale. Partial differential equations like (\ref{langevin}) have many applications. In cosmology, they can correspond to the Klein-Gordon equation in the Friedmann-Robertson-Walker metrics, where $\eta $ can be related to the Hubble parameter \cite{lg2}. In condensed matter physics, they describe a variety of phase transitions in systems such as superconductors and superfluids, including gaseous Bose-Einstein condensate.

Numerical simulations of Eq. (\ref{langevin}) can be computationally expensive, particularly for large systems and long quench timescales. This often limits the ability to study the defect formation in slow quenches.%Here, we introduce the following approach to overcome this difficulty.

\begin{figure}
{%
\includegraphics[clip,width=0.7\columnwidth]{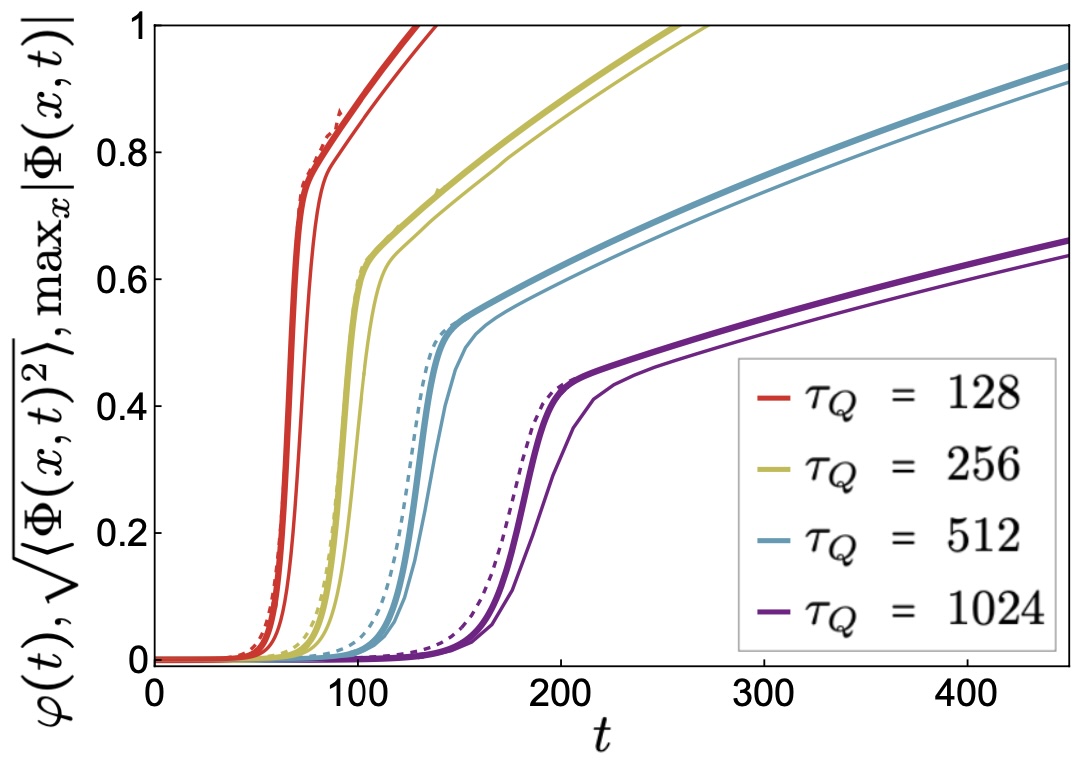} 
}
\caption{The thick lines represent the analytical solutions $\varphi (t)$ (Eq. (\ref{solu})) with $\eta=1$ and $\varphi (0)=10^{-4}$  for various quench timescales $\tau_Q$, while the numerical results $\sqrt{\langle \Phi (x,t)^2\rangle}$ (solid lines), $\displaystyle\mbox{max}_x |\Phi (x,t)|$ (dashed lines) are obtained by solving  Eq. (\ref{langevin}) with $\eta=1$ and $\theta=10^{-8}$. From left to right, $\tau_Q=128,256,512,1024$ respectively.} 
\label{fig2}
\end{figure}

\begin{figure}
{%
\includegraphics[clip,width=0.8\columnwidth]{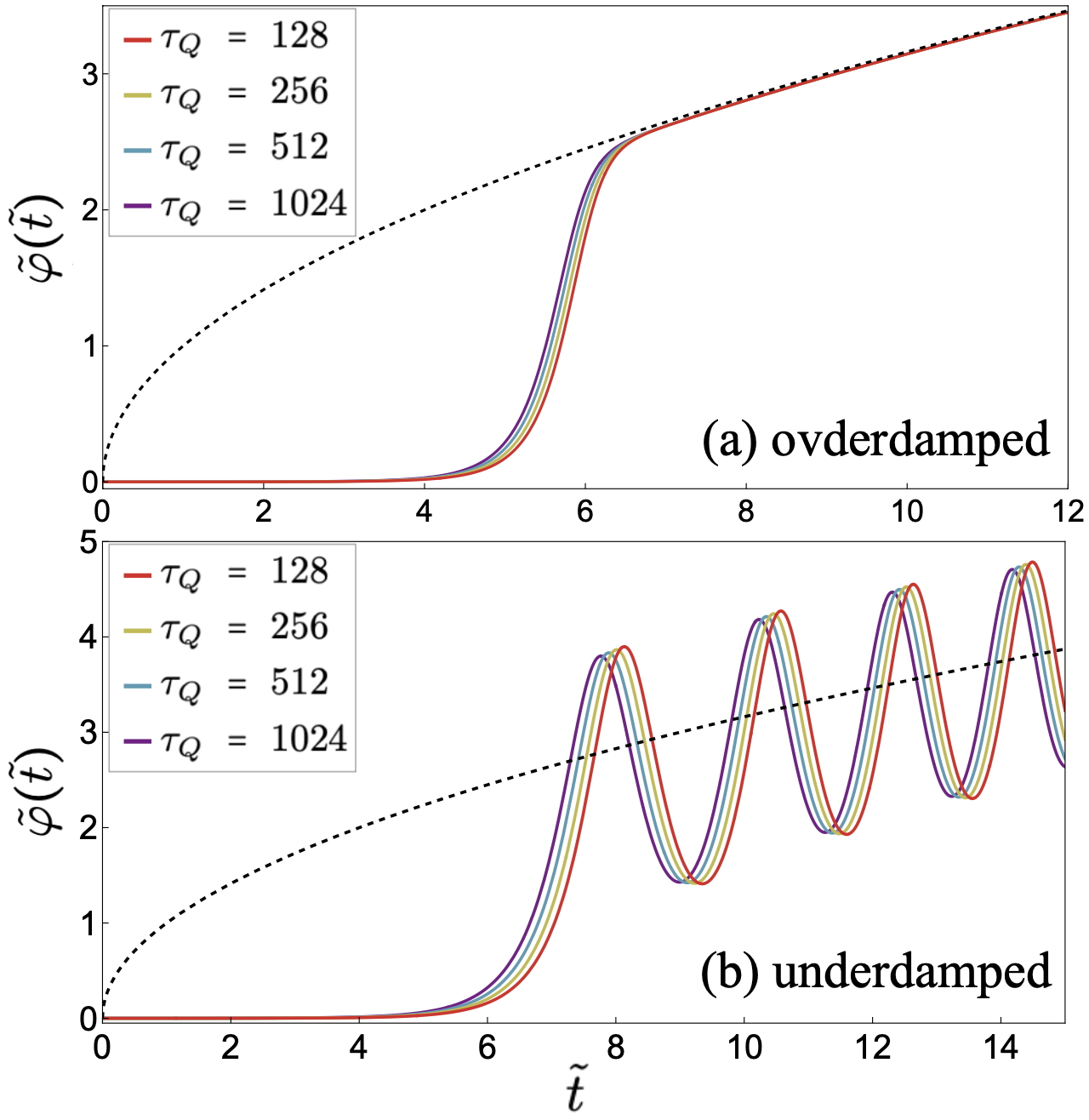} 
}
\caption{The rescaled  solution $\tilde{\varphi} (\tilde{t})$ in the overdamped case with $\eta=1$ where the first term of Eq. (\ref{ode1}) is discarded (a) and in the underdamped case  where the second term of Eq. (\ref{ode1}) is discarded (b).  $\varphi (0)=10^{-4}$ and various quench timescales $\tau_Q$. The dashed line represents $\tilde{\varphi}=\sqrt{\epsilon (\tilde{t})}$, the location of the minimum of the potential $V$.} 
\label{fig3}
\end{figure}

\begin{figure*}
{%
\includegraphics[clip,width=2\columnwidth]{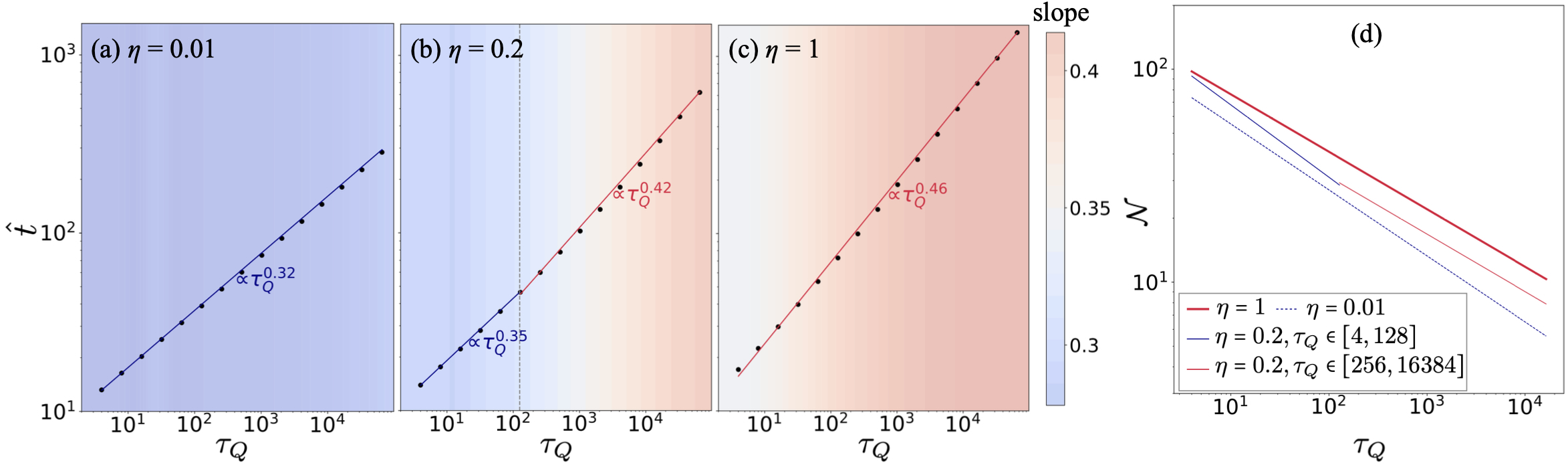} 
}
\caption{The freeze-out time $\hat{t}$ as a function of the quench timescale $\tau_Q$ for damping constants (a) $\eta = 0.01$, (b) $\eta = 0.2$, and (c) $\eta=1$, with $\varphi (0)=10^{-4}$. The color plot represents the slope of the log plot based on nearest neighbor points. For $\eta = 0.2$, a transition from the underdamped regime to the overdamped regime is observed as $\tau_Q $ increases. The dashed gray line represents the theoretical prediction  $\tau_Q = 1/\eta^3$ where the transition occurs. (d) The number of defects $\mathcal{N}$ as the function of $\tau_Q$ for $\eta=0.01$ (dashed blue line), $\eta=0.2$ (solid blue line for $\tau_Q\in[4,128]$ and solid red line for $\tau_Q\in[128,16384]$), and $\eta=1$ (thick red line) with $\theta=10^{-8}$.} 
\label{fig4}
\end{figure*}

Mindful of the paramount role of
% insensitivity of KZM to everything except 
critical slowing down, we consider a dramatic simplification of Eq. (\ref{langevin}) by omitting (for now) the spatial degrees of freedom and the noise term. As a result, we obtain an ordinary differential equation in which $\varphi (t)$ depends solely on time:
\begin{eqnarray}\label{ode1}
\ddot{\varphi}(t)+\eta \dot{\varphi}(t)=-\partial_{\varphi}V(\varphi(t)).
\end{eqnarray}
This  equation captures the time-dependent behavior of the order parameter $\varphi (t)$  without the numerical simulation of the full partial differential equation Eq. (\ref{langevin}). We now consider its properties and its relevance for KZM. We recognize that in the overdamped case where $\ddot{\varphi}(t) \ll \eta \dot{\varphi}(t)$, Eq. (\ref{ode1}) is a Bernoulli differential equation, 
\begin{eqnarray}
\eta \dot{\varphi}(t)+\frac12 (\varphi (t)^3-\epsilon (t) \varphi (t))=0
\end{eqnarray}
which can be solved analytically:
\begin{eqnarray}\label{solu}
\frac{\varphi (t)}{\varphi (0) }=\frac{ e^{t^2/4\eta\tau_Q}}{\sqrt{1+\varphi (0)^2\sqrt{\frac{\pi \tau_Q}{2\eta}}\mathrm{erfi}(\frac{t}{\sqrt{2\eta\tau_Q}})}}
\end{eqnarray}
where $\varphi (0) =\varphi (t=0)$ and $\mathrm{erfi}(y)=\frac{2}{\sqrt{\pi}}\int_0^y e^{y'^2}dy'$.

Fig. \ref{fig1} (a) shows the plot of the  solution (Eq. (\ref{solu})) with $\eta=1$, $\tau_Q=128$ for several values of $\varphi (0)$. All exhibit similar behavior: The initial period where $\varphi (t)$ slowly increases is followed by a ``jump" where the solution ``catches up" with the equilibrium value, and thereafter follows the equilibrium value of $\varphi (t)$ dictated by the broken symmetry minimum of the Landau-Ginzburg potential. This behavior is suggested by the adiabatic-impulse-adiabatic scenario \cite{whz,whz2,whz3}, and was seen in numerical simulations \cite{adolfo, jump}, as the start of rapid rise can be identified with $+\hat{t}$, when the order parameter evolution switches to catch up with the equilibrium value $\sim \sqrt{\epsilon(t)}$. We also note that before $t=0$ (when the critical point is transversed) $\varphi(t)$ ``jumps down'' to relatively small values, reaching them at the instant suggestive of $-\hat{t}$. In the impulse regime, $t\in [-\hat{t},\hat{t}]$, evolution is Gaussian and gradual, compared to these two ``jumps". Noise plays a key role. The preexisting values of $\varphi(t)$ for $t<-\hat t$ are forgotten: The noise, in effect, resets the initial $\varphi(0)$, as can be seen by comparing numerical solution with the analytic solution to Bernoulli equation in Fig. \ref{fig1} (a).

Fig. 1 (b) shows the plots of $\log(\varphi)$ for various $\varphi(0)$. Black circles indicate the freeze-out time $\hat{t}$ for each $\varphi(0)$, where $\frac{d^2\varphi}{dt^2} = 0$ signifying a ``jump" which we identify with $\hat t$. It demonstrates that the freeze-out time $\hat{t}$ only depends logarithmically on $\varphi(0)$.

Numerical simulation of the Langevin equation with full spatial dependence, Eq. (\ref{langevin}) in (1+1) dimensions, is in good agreement with the analytical solution of $\varphi (t)$ as shown in Fig. \ref{fig2}: $\sqrt{\langle \Phi (x,t)^2\rangle}$ (thin lines) obtained by averaging over the spatial dependence of $\Phi (x,t)^2$ is somewhat smaller than the solution of Bernoulli equation given by Eq. (\ref{solu}) (thick lines). Its value is suppressed by the presence of topological defects. Maximum values of $|\Phi (x,t)|$, $\displaystyle\max_x |\Phi(x,t)|$ (dashed lines), by contrast, are slightly larger than $\varphi (t)$, enhanced by the random walk due to noise. This demonstrates that, to a large extent, the time evolution of the order parameter $\varphi$ can be effectively captured solely by solving the ordinary differential equation. The correspondence between $\varphi (0) $ and $\theta $ can be described as follows. Before $t=0$, the potential can be approximated by a harmonic potential $V_{\rm har} (\Phi)=-\frac{1}{4}\epsilon (t) \Phi^2$ near $\Phi = 0$.  Since the temperature $\theta$ corresponds to the kinetic energy of $\Phi$, we have $\sqrt{\langle \Phi^2\rangle}\approx \sqrt{\theta/(-2\epsilon (t))}$.  Order parameter $\Phi$ is subject to noise once it enters this regime. While perturbations present at $-\hat t$ would survive till $+\hat t$, noise will add to them and can be amplified. We can estimate effective $\varphi (0)$  as $\varphi (0) \approx  \sqrt{\theta/2\hat{\epsilon}}$ where $\hat{\epsilon}=\epsilon (\hat{t})$ at freeze-out time $\hat{t}$. Because $\hat{\epsilon}$ depends on the quench timescale $\tau_Q$, $\varphi (0)$ typically needs to be adjusted accordingly. However, for sufficiently small temperature $\theta$, this dependence can be ignored.

Plotting the rescaled solution $\tilde{\varphi}(\tilde{t})$ where $\tilde{\varphi}=\sqrt[4]{\tau_Q/\eta}\varphi$ and $\tilde{t}= t/\sqrt{\eta\tau_Q}$
  reveals kinship between the curves for different quench timescale $\tau_Q$ in the overdamped regime (Fig. \ref{fig3} (a)) and only a slow (logarithmic) dependence on the value of the order parameter $\varphi(0)$. 
  %Therefore, it can be readily anticipated 
  It follows that the freeze-out time $\hat{t}$ obeys the relationship \cite{whz, lg2}
\begin{eqnarray}\label{that1}
\hat{t}\propto \sqrt{\eta\tau_Q}
\end{eqnarray}
in the overdamped case. 

For the underdamped case, Eq. (\ref{ode1}) can be  solved numerically. Fig. \ref{fig3} (b) depicts the rescaled solution $\tilde{\varphi}(\tilde{t})$ of Eq. (\ref{ode1}) where the second term $\eta\dot{\varphi}$ is neglected and $\tilde{\varphi}=\tau_Q^{1/3}\varphi$, $\tilde{t}= t/\tau_Q^{1/3}$. After this rescaling, the solutions corresponding to different quench timescales $\tau_Q$ exhibit a close relationship .
%and only a slow (logarithmic) dependence on the value of the order parameter $\varphi(0)$.  
%Based on this observation, one can anticipate that t
Therefore, the freeze-out time $\hat{t}$ obeys the relationship \cite{whz,lg2}:
\begin{eqnarray}\label{that2}
\hat{t} \propto \tau_Q^{1/3}
\end{eqnarray}
in the underdamped case.

As previously discussed, by selecting the appropriate initial condition $\varphi (0) \approx  \sqrt{\theta/2\hat{\epsilon}}$, closer alignment among plots for each value of $\tau_Q$ in Fig. \ref{fig3} can be achieved in both the overdamped and underdamped cases. 
%However, for sufficiently small $\varphi (0)$, the logarithmic dependence of $\varphi (0)$ on $\hat{\epsilon}$ can be disregarded.

The scaling of the freeze-out time $\hat{t}$ can also be determined directly from  the solution of Eq. (\ref{ode1}). Analyzing the solution of the ordinary differential equation without assuming either the overdamped or underdamped limits reveals the transition from the scaling behavior of the overdamped regime to that of the underdamped regime. The freeze-out time $\hat{t}$ is defined as the moment when the solution $\varphi(t)$ rapidly begins to move toward the potential minima after $t=0$. This time corresponds to the point at which the second time-derivative $\frac{d^2 \varphi}{dt^2} =0$ for the first time after $t=0$. In Fig. \ref{fig4}, $\hat{t}$  as a function of the quench timescale $\tau_Q$ is plotted for (a) $\eta=0.01$, (b) $\eta=0.2$, and (c) $\eta=1$, with $\varphi (0)=10^{-4}$  obtained using this method. The color plot represents the slope of the log plot based on nearest neighbor points. It is found that $\hat{t}\propto \tau_Q^{0.46}$ for $\eta=1$ and 
$\hat{t}\propto \tau_Q^{0.32}$ for $\eta=0.01$, both of which closely align with the relations given by Eqs. (\ref{that1},\ref{that2}) respectively, as shown in \cite{lg2}. The color plot reveals a transition from the underdamped regime to the overdamped regime as $\tau_Q$ increases for $\eta=0.2$. We have that $\hat{t}\propto \tau_Q^{0.35}$ for  $\tau_Q\in [4,128]$ and $\hat{t}\propto \tau_Q^{0.42}$ for  $\tau_Q\in [128, 16384]$. The gray dashed line represents the theoretical prediction $\tau_Q=1/\eta^3$, where the transition occurs \cite{lg2, suppl}. Fig. \ref{fig4} (d) shows the number of defects $\mathcal{N}$ created by phase transitions as a function of $\tau_Q$. This result is obtained by numerically solving the full partial differential equation (Eq. (\ref{langevin})) in (1+1) dimensions 15 times, starting the time evolution at  $\epsilon=-1$ and concluding it at 
$t=32768$ (i.e., $\epsilon=1$ for $\tau_Q=16384$). The system size is $L=2048$ with 4096 grid points, and $\theta=10^{-8}$. In the underdamped case with $\eta=0.01$, the number of defects $\mathcal{N}\propto \tau_Q^{-0.31}$ (dashed blue line), while $\mathcal{N}\propto \tau_Q^{-0.27}$ in the overdamped case with $\eta=1$ (thick red line). For $\eta=0.2$, we observe a transition in the scaling behavior of the number of defects. We have $\mathcal{N}\propto \tau_Q^{-0.34}$ for small quench timescales $\tau_Q \in [4,128]$ and $\mathcal{N} \propto \tau_Q^{-0.27}$ for large quench timescales $\tau_Q\in [128,16384]$.
 It can be observed that the scaling of the number of defects clearly reflects the change in the scaling of the freeze-out time $\hat{t}$ from the underdamped to the overdamped regime as $\tau_Q$ increases. This behavior \cite{whz, lg2} can be predicted solely by solving the ordinary differential equation (Eq. (\ref{ode1})).\\
 
 \begin{figure}
{%
\includegraphics[clip,width=0.9\columnwidth]{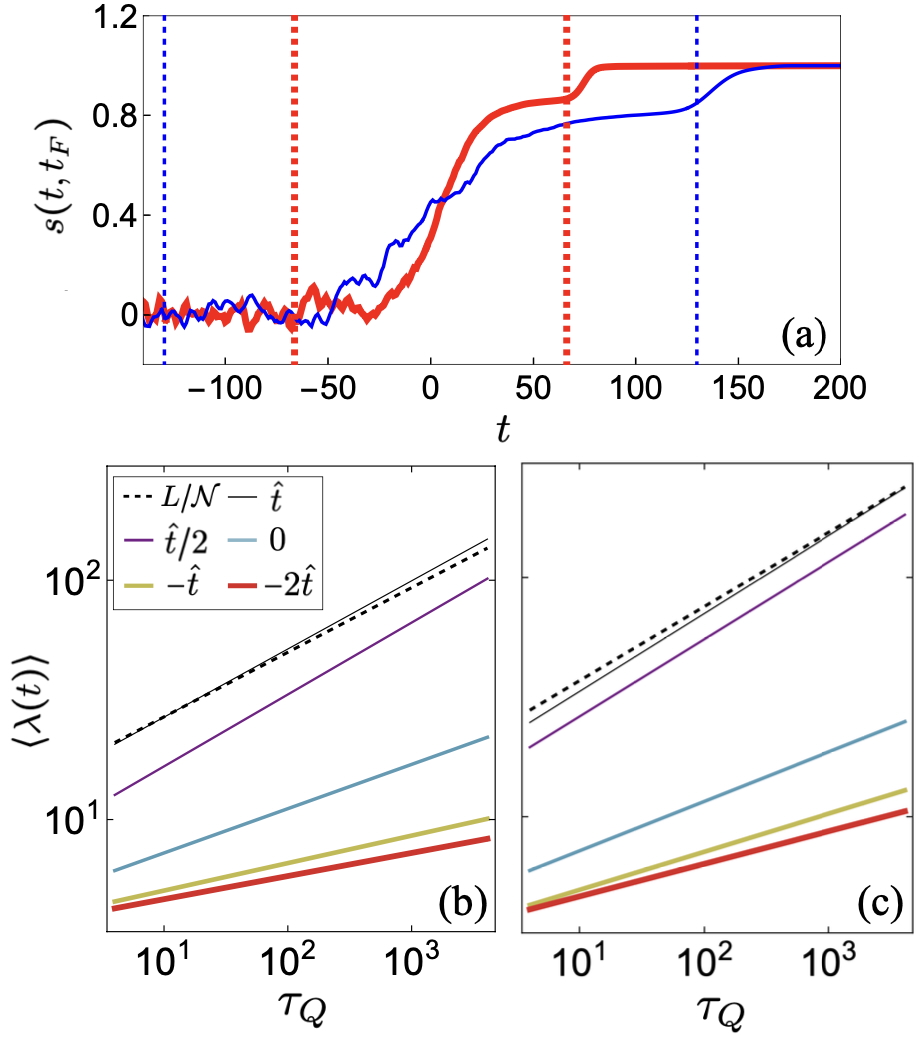} 
}
\caption{ (a) $s(t,t_F)$ for $\tau_Q=128$ (thick, red) and $\tau_Q=512$ (blue). $t_F=300$ and $\theta=10^{-8}$. Vertical dashed lines indicate $\pm \hat{t}$. (b,c) $\langle \lambda (t) \rangle $ as a function of $\tau_Q$ for   $\eta=1$ (b) and for  $\eta=0.01$ (c). From thick red to thin black line, $t=-2\hat{t}, -\hat{t}, 0, \hat{t}/2, \hat{t}$ respectively. The dashed black lines represent $L/\mathcal{N}$ (i.e., the average domain size) from Fig. \ref{fig4} (d). $\theta=10^{-8}$.} 
\label{fig5}
\end{figure}

We now turn to the question: How are the timescales ($\hat{t}$) imprinted on the spatial structure of $\Phi(x,t)$?  To address this question, we consider the equation that excludes the time dependence and the noise term from Eq. (\ref{langevin}). The density of defects results from the interplay of the temporal evolution of the order parameter and the dependence $\phi $ on $x$. The equation for $\phi $:
\begin{eqnarray}\label{spatial}
\nabla^2 \phi (\mathbf{x}) -\partial_{\phi} V(\phi) =0
\end{eqnarray}
is solved as
\begin{eqnarray}\label{jacobi}
\phi (x)=\phi(0)\mbox{cd}\left(\frac{i\sqrt{\phi(0)^2-2\epsilon}x}{2},-\frac{\phi(0)^2}{\phi(0)^2-2\epsilon}\right)
\end{eqnarray}
in 1-dimensional space where $\phi(0)=\phi(x=0)$, $\phi'(0)=0$ and $\mbox{cd}$ represents Jacobi elliptic function. Eq. (\ref{jacobi}) can be approximated by $\phi (x)=\phi (0) \cos \sqrt{\epsilon/2}x$ for small $|\phi (0)| <<\sqrt{\epsilon}$ when $\epsilon >0$. This exhibits spatial periodicities related to $1/ \sqrt \epsilon $. We are particularly interested in periodicity $1/ \sqrt {\hat \epsilon} $ exhibited for small $\phi(0)$: At about $+\hat t$ order parameter $\Phi(x,t)$ begins to grow from the small pre values bestowed during the $[- \hat t, +\hat t]$ interval  by the random walk $\theta(x,t)$ in the Langevin equation Eq. (\ref{langevin}), rising to the broken symmetry equilibrium $\sqrt{\epsilon}$. Focusing on its spatial part, we can think of the local $\phi(x)$ as a consequence of the random walk filtered by the spatial component of the Langevin equation. That filtering imposes periodicities (set by the ``spring constant’’ $\sim  \epsilon $) of the Jacobi function (which represents the solution of the ``physical pendulum’’. The period is exactly $ \sim 1/\sqrt \epsilon $ when, for small values of $\phi$ the effect of the nonlinearity in Eq. (\ref{spatial}) can be neglected. This yields
\begin{eqnarray}
\hat{\xi} \sim   \begin{cases}
                        (\tau_Q/\eta)^{1/4}, \qquad \text{  (overdamped case)} \\
                        \tau_Q^{1/3}, \qquad \qquad \quad \text{(underdamped case)}
                    \end{cases}.
\end{eqnarray}

That structure is both imprinted and erased on $\Phi (x,t)$ by the combination of random walk due to the noise term and the dynamics in the interval $[- \hat t, +\hat t]$. The evidence of its gradual accumulation (that at $t > +\hat t$ leads to the formation of topological defects) can be seen in the time-dependent scalar product:
\begin{eqnarray}
 s(t,t_F) = \frac{\sum_{x}\Phi (x,t)\Phi (x,t_F)}{\sqrt{\sum_{x}\Phi (x,t)^2}\sqrt{\sum_x \Phi (x,t_F)^2}}
\end{eqnarray}
where $t_F$ is the time after $\hat{t}$. In Fig. \ref{fig5} (a), we see that $s(t,t_F)$ begins to rise already before the critical point is traversed, soon after $-\hat t$, and reaches its equilibrium value shortly after $+ \hat t$ when, according to \cite{whz}, the basic structure of the broken symmetry state (including the location of the defects) is determined. This confirms the adiabatic-impulse-adiabatic parsing of the dynamics of symmetry breaking quenches. 

Finally, we examine the time evolution of the expectation of half the spatial period $\langle \lambda (t)\rangle$ given by the Fourier transform, defined as follows:
\begin{eqnarray}
\frac{1}{\langle \lambda (t)\rangle}=\frac{1}{\mathcal{C}}\sum_k \frac{2k}{L} \tilde{\Phi} (k,t) \tilde{\Phi}^{*} (k,t)
\end{eqnarray}
where $ \tilde{\Phi} (k,t)=\sum_{x}\Phi (x,t) e^{-i 2\pi  k x/L}$, the normalisation $\mathcal{C}= \sum_k \tilde{\Phi} (k,t) \tilde{\Phi}^{*} (k,t)$, and $L=2048$ is the system size. Here, we consider  half the spatial period $L/2k$ since the full period of the oscillation would produce a pair of kinks. Fig. \ref{fig5} (b,c) shows $\langle \lambda (t) \rangle $ as a function of $\tau_Q$ from $t=-2\hat{t}$ to $\hat{t}$ for the overdamped case with $\eta=1$ (b) and for the underdamped case with $\eta=0.01$ (c). The figure is obtained by averaging the results of 15 numerical simulations of Eq. (\ref{langevin}). As time approaches $\hat{t}$, the behavior governed by the periodicity from the spatial component of the Langevin equation becomes dominant over that driven by the random walk. Finally, the scaling of $\langle \lambda (t)\rangle$ aligns closely with $L/\mathcal{N}$ (i.e., the average domain size) at $t=\hat{t}$ where $\mathcal{N}$ is the number of defects from Fig. \ref{fig4} (d).\\

We demonstrated that solving the ordinary differential equations governing the temporal or spatial components of the Langevin equation yields surprisingly valuable insights into the KZM. These results provide a deep understanding of topological defect formation in second-order phase transitions by simply solving ordinary differential equations,  and enable the study of Kibble-Zurek scaling over a large range   of quench timescales. The supplemental material \cite{suppl} provides a discussion of the correlation length using the Ornstein-Zernike form, as well as an analysis of freeze-out time and correlation length for general potentials using the approach presented in this paper. There is no reason to believe that KZM scaling would yield correct prediction for defect density in such exotic potential, although preliminary study suggests that this is the case as long as the exponents are not too different form the Landau-Ginzburg potential, but it is no longer accurate when these exponents are substantially different.

\begin{acknowledgements}
We thank John Bowlan for helpful discussions. F.S. acknowledges support from the Los Alamos National Laboratory LDRD program under project number 20230049DR and the Center for Nonlinear Studies under project number 20250614CR-NLS.
\end{acknowledgements}

\bibliographystyle{apsrev4-1}
\bibliography{test} 
\appendix
\onecolumngrid
\renewcommand{\theequation}{A.\arabic{equation}}
\setcounter{equation}{0}
\section*{Supplemental material}

\subsection{Correlation length}

In addition to the method used in the main text, we can also estimate the correlation length as follows.   We consider  the following equation that excludes the time dependence and the noise term from Eq. (\ref{langevin}), and by introducing a delta-function source $-\delta (x)$ at the origin:
\begin{eqnarray}
\nabla^2 \phi (\mathbf{x}) -\partial_{\phi} V(\phi) =-\delta (\mathbf{x}).
\end{eqnarray}
We assume $\phi (\mathbf{x})=\phi_0 (\mathbf{x}) +\delta \phi (\mathbf{x})$ where $\nabla^2 \phi_0 (\mathbf{x}) -\partial_{\phi_0} V(\phi_0)=0$ while the perturbation $\delta \phi$ obeys
\begin{eqnarray}
\nabla^2 \delta \phi (\mathbf{x})+\frac12  \epsilon (t)\delta \phi (\mathbf{x})=-\delta (\mathbf{x}).
\end{eqnarray}
Here the higher-order terms of $\delta \phi (\mathbf{x})$ are discarded. The solution of this equation takes the Ornstein-Zernike form:
\begin{eqnarray}
\delta \phi (\mathbf{x}) \sim |\mathbf{x}|^{-(d-1)/2}\exp (-|\mathbf{x}|/\xi)
\end{eqnarray}
in $d$-dimensional space and $\xi \sim 1/\sqrt{\epsilon}$. This yields
\begin{eqnarray}
\hat{\xi} \sim   \begin{cases}
                        (\tau_Q/\eta)^{1/4}, \qquad \text{  (overdamped case)} \\
                        \tau_Q^{1/3}, \qquad \qquad \quad \text{(underdamped case)}
                    \end{cases}.
\end{eqnarray}

\subsection{General potential}

Although it remains ambiguous whether the Kibble-Zurek mechanism  is applicable to general potentials, we demonstrate here that the scaling of the freeze-out time and correlation length for a general potential can be estimated using the methodology presented in the main text. We consider a general potential: 
\begin{eqnarray}\label{pot}
V(\Phi)=(\Phi^{2m}-2\epsilon (t) \Phi^{2n})/8
\end{eqnarray}
where $m>n$. In the main text, the Landau-Ginzburg case where $m=2$ and $n=1$ has been studied. The behavior of $\hat{t}$ for this general potential can  be examined as follows.

As evident from Fig. \ref{fig3} in the main text, it is useful to identify the rescaling of $\varphi (t)$ and $t$ that makes the solution independent of the quench timescale $\tau_Q$, as this reveals the scaling behavior of the freeze-out time $\hat{t}$. This can be achieved by making the ordinary differential equation  independent  of the damping constant $\eta$ and $\tau_Q$. 

For the overdamped case, the ordinary differential equation with a general potential $V(\varphi)=(\varphi^{2m}-2\epsilon (t)\varphi^{2n})/8$ can be written as
\begin{eqnarray}
\eta \frac{d\varphi}{dt}+\frac14 (m\varphi^{2m-1}-2n \epsilon \varphi^{2n-1})=0.
\end{eqnarray}

 By applying the transformation $\varphi=\eta^{a}\tau_Q^{a'}\tilde{\varphi}$ and $t=\eta^{b}\tau_Q^{b'}\tilde{t}$, we obtain 
\begin{eqnarray}
\eta^{a-b+1}\tau_Q^{a'-b'}\frac{d\tilde{\varphi}}{d\tilde{t}}+\frac14 (m\eta^{(2m-1)a}\tau_Q^{(2m-1)a'}\tilde{\varphi}^{2m-1}-2n \tilde{t} \eta^{(2n-1)a+b}\tau_Q^{(2n-1)a'+b'-1}\tilde{\varphi}^{2n-1})=0.
\end{eqnarray}

This equation reduces to
\begin{eqnarray}
\frac{d\tilde{\varphi}}{d\tilde{t}}+\frac14 (m \tilde{\varphi}^{2m-1}-2n \tilde{t}\tilde{\varphi}^{2n-1})=0
\end{eqnarray}
which is independent from $\eta$ and $\tau_Q$ when 
\begin{eqnarray}
a&=&\frac{1}{2(2m-n-1)}, \quad a'=\frac{1}{2(n-2m+1)}\nonumber\\
b&=&\frac{m-n}{2m-n-1}, \quad b'=\frac{m-1}{2m-n-1}.
\end{eqnarray}
Therefore,
\begin{eqnarray}
\hat{t}\propto \eta^{(m-n)/(2m-n-1)}\tau_Q^{(m-1)/(2m-n-1)}
\end{eqnarray}
in the overdamped regime.

For the underdamped regime, we have
\begin{eqnarray}
\eta^{a-2b}\tau_Q^{a'-2b'}\frac{d^2 \tilde{\varphi}}{d\tilde{t}^2}+\frac14 (m\eta^{(2m-1)a}\tau_Q^{(2m-1)a'}\tilde{\varphi}^{2m-1}-2 n \tilde{t} \eta^{(2n-1)a+b}\tau_Q^{(2n-1)a'+b'-1}\tilde{\varphi}^{2n-1})=0.
\end{eqnarray}

This equation reduces to
\begin{eqnarray}
\frac{d^2\tilde{\varphi}}{d\tilde{t}^2}+\frac14 (m\tilde{\varphi}^{2m-1}-2n \tilde{t}\tilde{\varphi}^{2n-1})=0
\end{eqnarray}
which is independent from $\tau_Q$ when
\begin{eqnarray}
a&=&0, \quad a'=\frac{1}{2n-3m+1},\nonumber\\
b&=&0, \quad b'=\frac{m-1}{3m-2n-1},
\end{eqnarray}
and we obtain
\begin{eqnarray}
\hat{t} \propto \tau_Q^{(m-1)/(3m-2n-1)}.
\end{eqnarray}

In particular, we find that $\hat{t}\propto \sqrt{\eta\tau_Q}$ for the overdamped case and  $\hat{t}\propto \tau_Q^{1/3}$ for the underdamped case when $n=1$.

The overdamped scalings are applicable when the evolution is governed by the first derivative term (i.e., $\eta \dot{\varphi}>\ddot{\varphi}$) at the moment topological defects freeze out. This occurs when the freeze-out time $\hat{t}$ for the overdamped case is larger than that for the underdamped case, which corresponds to
\begin{eqnarray}
\tau_Q>\eta^{-(3m-2n-1)/(m-1)}.
\end{eqnarray}
Therefore, it was estimated that the transition from the underdamped regime to the overdamped regime occurs around $\tau_Q\sim 1/\eta^3$ for $m=2$ and $n=1$  in Fig. \ref{fig4} (b) of the main text.\\

The scaling of the correlation length $\xi$ can also be found similarly. We consider the ordinary differential equation for the spatial dependence $\phi$:
\begin{eqnarray}
\frac{d^2 \phi (x)}{dx^2}+ \frac{n}{2} \epsilon \phi(x)^{2n-1}  =0.
\end{eqnarray}
Here, we neglect the higher-order term associated with $m$.

By applying the transformation $x=\epsilon^{-1/2} \tilde{x}$, we obtain the equation that becomes independent of $\epsilon$, and we can anticipate that $\xi \sim 1/\sqrt{\epsilon}$.

However,  KZM-like scaling can cease to be applicable when $n$ differs significantly from $n=1$. This appears to depend on the noise temperature. It is not surprising, since critical slowing down is key for KZM.

\end{document}